\documentclass[aps,prl, notitlepage, twocolumn,superscriptaddress,floatfix,letterpaper,nofootinbib]{revtex4-1}

\usepackage{amsmath, amsthm, amssymb,slashed}

\usepackage[usenames, dvipsnames]{color}
\usepackage[svgnames]{xcolor}
\usepackage[colorlinks,citecolor=RoyalBlue, urlcolor=RoyalBlue, linkcolor=RoyalBlue ]{hyperref} 

\usepackage[normalem]{ulem}

\usepackage{booktabs}

\definecolor{mygray}{gray}{0.6}

\usepackage{upgreek}
\usepackage{bbm}

\newenvironment{myfont}[2][]{\csname#2\endcsname[#1]}{}

\usepackage{slashed}
\usepackage[makeroom]{cancel}
\usepackage[normalem]{ulem}
\usepackage{soul}
\newcommand{\stkout}[1]{\ifmmode\text{\sout{\ensuremath{#1}}}\else\sout{#1}\fi}

\usepackage{sseq}
\usepackage[all,cmtip]{xy}
\usepackage{tikz-cd}
\usepackage{tikz}
\usetikzlibrary{matrix}
\usetikzlibrary{decorations.markings}
\usetikzlibrary{tikzmark,decorations.pathreplacing,positioning}
\usepackage{amsfonts}

\newcommand{\bea}{\begin{eqnarray}}
\newcommand{\eea}{\end{eqnarray}}
\def\be{\begin{equation}}
\def\ee{\end{equation}}

\newcommand{\e}{\hspace{1pt}\mathrm{e}}

\newcommand{\ii}{\hspace{1pt}\mathrm{i}\hspace{1pt}}

\definecolor{red}{rgb}{1,0,0}
\definecolor{blue}{rgb}{0,0,1}
\definecolor{dblue}{rgb}{0,0,0.4}
\definecolor{green}{rgb}{0,1,0}
\definecolor{black}{rgb}{0,0,0}
\definecolor{white}{rgb}{1,1,1}

\definecolor{brn}{rgb}{.8,.4,.0}
\definecolor{redo}{rgb}{1,.5,.0}
\definecolor{ddgrn}{rgb}{0,0.4,0}
\definecolor{dgrn}{rgb}{0,0.55,0}
\definecolor{dbl}{rgb}{0,0,0.5}

\usepackage[bbgreekl]{mathbbol}
\usepackage{amscd}

\newcommand{\Z}{\mathbb{Z}}

\newcommand{\R}{\mathbb{R}}

\newcommand{\dd}{\hspace{1pt}\mathrm{d}}
\newcommand{\<}{\langle} 
\renewcommand{\>}{\rangle} 

\newcommand{\Refe}[1]{Ref.~[\onlinecite{#1}]}
\newcommand{\Eq}[1]{Eq.~(\ref{#1})} 
\newcommand{\eq}[1]{eq.~(\ref{#1})}

\newcommand{\Tr}{{\rm Tr}}

\newcommand{\prt}{\partial}

\newcommand{\up}{\uparrow} 
\newcommand{\down}{\downarrow}

\newcommand{\bpm}{\begin{pmatrix}}
\newcommand{\epm}{\end{pmatrix}}
\newcommand{\bmm}{\begin{matrix}}
\newcommand{\emm}{\end{matrix}}

\newcommand{\cH}{ {\cal H} }

\newcommand{\al}{\alpha} 
\newcommand{\bt}{\beta}

\def\Z{{\mathbb{Z}}}

\def\R{{\mathbb{R}}}

\def\Tr{{\mathrm{Tr}}}

\def \Z{\mathbb{Z}}

\newcommand {\emptycomment}[1]{}

\newcommand{\SO}{{\rm SO}}
\newcommand{\Spin}{{\rm Spin}}
\newcommand{\U}{{\rm U}}
\newcommand{\SU}{{\rm SU}}

\newcommand{\Pin}{{\rm Pin}}

\newcommand{\rO}{{\rm O}}

\usepackage{centernot}
\newcommand{\nn}{{\nonumber}}
\usepackage{enumitem} 
\usepackage{mathtools,amssymb,varwidth}

\newcommand{\Fig}[1]{Fig.~\ref{#1}} 
\newcommand{\Table}[1]{Table \ref{#1}}

\usepackage{datetime}

\def\bM{{\mathbb{M}}}
\def\bD{{\mathbb{D}}}
\def\bQ{{\mathbb{Q}}}

\newcommand{\rC}{{\rm C}}
\newcommand{\rP}{{\rm P}}
\newcommand{\rT}{{\rm T}}
\newcommand{\rR}{{\rm R}}
\newcommand{\rF}{{\rm F}}
\newcommand{\rB}{{\rm B}}
\newcommand{\rK}{{\rm K}}

\begin{document}


\title{C-P-T Fractionalization}

\author{Juven Wang 
}
\email{jw@cmsa.fas.harvard.edu}
\homepage{http://sns.ias.edu/~juven/}
\affiliation{Center of Mathematical Sciences and Applications, Harvard University, MA 02138, USA}

\begin{abstract} 
Discrete spacetime symmetries of parity P or reflection R, and time-reversal T, act naively as  
$\mathbb{Z}_2$-involutions in the \emph{passive} transformation on the spacetime coordinates;
but together with a charge conjugation C,
the total C-P-R-T symmetries have enriched \emph{active} transformations 
on fields in representations of the spacetime-internal symmetry groups of quantum field theories (QFTs).
In this work, we derive 
that 
these symmetries can be further fractionalized, especially in the presence of the fermion parity $(-1)^{\rm F}$.
We elaborate on examples including
relativistic Lorentz invariant QFTs (e.g., 
spin-1/2 Dirac or Majorana spinor fermion theories)
and nonrelativistic quantum many-body systems (involving Majorana zero modes),
and comment on applications to spin-1 Maxwell electromagnetism (QED) or interacting Yang-Mills (QCD) gauge theories.
We discover various C-P-R-T-$(-1)^{\rm F}$ group structures,
e.g., 
Dirac spinor is in a \emph{projective} representation of $\mathbb{Z}_2^{\rm C}\times \mathbb{Z}_2^{\rm P} \times \mathbb{Z}_2^{\rm T}$ 
but in an (\emph{anti})\emph{linear} representation of
an order-16 nonabelian finite group, 
as the central product between an order-8 dihedral (generated by C and P) or quaternion group 
and an order-4 group generated by T with T$^2=(-1)^{\rm F}$. 
The general theme may be coined as C-P-T or C-R-T fractionalization.




\end{abstract}


\maketitle


\section{Introduction and Summary}

Common physics knowledge recites
that the time-reversal T and parity P
are \emph{discrete} spacetime symmetries that cannot be continuously deformed from the identity element ---
T and P are not part of the proper orthochronous restricted \emph{continuous} Lorentz symmetry group  
$\SO^+(d,1)$. 
It is important to distinguish the T and P from the mirror reflection R.
As \emph{passive} transformations on the spacetime coordinates $x \equiv (t,\vec{x})$, 
\bea \label{eq:coordinate}
 \rT (t,{x_1}, \dots, {x_d}) \rT^{-1} &=&  x'_{\rT} \equiv (-t,{x_1}, \dots, {x_d}),\nn\\
 \rP (t,{x_1}, \dots, {x_d}) \rP^{-1}&=& x'_{\rP} \equiv (t,-{x_1}, -\dots, -{x_d}),\\
\rR (t,{x_1}, \dots, {x_d}) \rR^{-1}&=& x'_{\rR} \equiv (t,-{x_1}, +\dots, +{x_d}),\nn
\eea
where T flips the time coordinate, 
P flips all $\vec{x}$, but
R flips only on one coordinate (here say $x_1$) with respect to a mirror plane (normal to $x_1$).
We label the spacetime coordinate 
component $x_\mu$ with $\mu=0,1,\dots,d$ for $(d+1)$-spacetime dimensions (denoted as $d+1$d). 
The transformed coordinates are labeled as $x'$, or $x'_\mu$ for each component,
with the subscript T/P/R/$etc.$ to indicate which coordinates are transformed.
In odd-dimensional spacetime, the P is in fact a subgroup of a continuous spatial rotational symmetry special orthogonal 
$\SO(d) \subset \SO^+(d,1)$, 
thus unluckily P is not an independent discrete symmetry.
We should replace P by the reflection R. For example, the 
CPT theorem \cite{SchwingerPhysRev.82.914, Pauli1955, Pauli1957, Luders1954, LUDERS19571, StreaterWightman1989} 
should be called the CRT theorem \cite{Witten1508.04715, Freed1604.06527} in any general dimension of Minkowski spacetime.
In this work, we mainly focus on the even-dimensional spacetime, so we can choose either P or R symmetry. 
We shall mainly use P to match the major literature, but we will comment about R when necessary. 

Charge conjugation C, however, cannot manifest itself under a \emph{passive} transformation on the spacetime coordinates,
but can reveal itself under an \emph{active} transformation on a particle or field, 
such as a complex-valued spin-0 Lorentz scalar $\phi(x)=\phi(t,\vec{x})$ (which is a function of the spacetime coordinates).
The C colloquially flips between particle and anti-particle sectors,
or more generally between energetic excitations and anti-excitations
\be
\rC \text{(excitations)} \rC^{-1} =  \text{(anti-excitations)}
\ee
involving the complex conjugation (denoted ${}^*$).
The \emph{active} transformation acts on this Lorentz scalar $\phi$ as
\bea
\rC \phi(t,\vec{x}) \rC^{-1} &=&\phi'_\rC(t,\vec{x})=\phi^*(t,\vec{x})=\phi^*(x),\nn\\
\rP \phi(t,\vec{x}) \rP^{-1} &=&\phi'_\rP(t,\vec{x})= \phi(t,-\vec{x})=\phi(x'_{\rP}),\\ 
\rT \phi(t,\vec{x}) \rT^{-1} &=&\phi'_\rT(t,\vec{x})=\phi(-t,\vec{x})=\phi(x'_{\rT}).\nn
\eea

All the above transformations, regardless \emph{passive} or \emph{active}, naively seem to be only $\Z_2$-involutions in mathematics,
such that twice transformations are the null (do nothing) 
transformations.\footnote{Let us 
clarify the \emph{passive} vs \emph{active} transformations, and their involution.
Suppose we take a spatial coordinate $x$ and a scalar function $f(x)$ as an example,
the \emph{passive} transformation $F_p$ maps $(x,f(x))$ to $(-x,f(x))$,
while the \emph{active} transformation $F_a$ maps $(x,f(x))$ to $(x,f(-x))$.
So we see that both $F_p(F_p(x,f(x)))=(x,f(x))$
and $F_a(F_a(x,f(x)))=(x,f(x))$ are $\Z_2$-involutions such that $F_p$ and $F_a$ are their own inverse functions.
The above discussion also follows for the time coordinate $t$, by replacing $x$ with $t$.
However, we will take the \emph{active} transformation viewpoint on the classical fields or quantum fields.
We shall 
reveal their fractionalization of C-P-R-T symmetries, beyond this $\Z_2$-involution structure.
}
Thus it reveals a finite group of order 2 structure, namely $\Z_2$.

In this scalar field
example, the C-P-T symmetry form a direct product group
$\mathbb{Z}_2^C \times \mathbb{Z}_2^P \times \mathbb{Z}_2^T$. 
One may mistakenly conclude $\rC^2=\rP^2=\rR^2=\rT^2=+1$ and assume they are all commute in general.
The essence of our work is to point out that
all these ``discrete C, P, R, or T symmetries''
(which we denote altogether as ``C-P-R-T'' in short)
can form a rich nonabelian finite group structure, in the physical realistic systems pertinent to experiments or theories. 
%
%
We can possibly fractionalize the C-P-R-T group structures further,
for the statevectors in quantum mechanics
or the fields in classical or quantum field theories (QFTs), in various representations ({rep}) of the spacetime or internal symmetry groups 
(denoted as $G_{\text{spacetime}}$ and $G_{\text{internal}}$).

The {\bf \emph{symmetry fractionalization}} 
means the following: the matter field is \emph{not} in the \emph{linear representation} of the original symmetry group $G$,
but in the \emph{projective representation} of $G$ and 
in the \emph{linear representation} of the extended total group $\tilde G$. 
A typical case is illustrated by a group extension 
$
1 \to N \to \tilde G \to G \to 1
$
where $G$ is the quotient group  
while the $N$ is the normal subgroup of the total group $\tilde G$, so $\tilde G/N=G$.
A famous example is the gapped
1+1d isospin-1 Haldane chain with $G=\SO(3)$ symmetry \cite{Affleck1988nt1989QuantumSpinChainsHaldaneGap},
whose 0+1d boundary can host a two-fold degenerated 
isospin-1/2 doublet of $\tilde G= \SU(2)$, with $N=\Z_2$.
Thus this doublet is in a projective rep of $G=\SO(3)$, also in a linear rep $\tilde G= \SU(2)$.

In this work, we will find the analogous {\bf C-P-T \emph{symmetry fractionalization}}.
For example, in contrast to a spin-0 scalar field's $G_{\phi} \equiv \mathbb{Z}_2^C \times \mathbb{Z}_2^P \times \mathbb{Z}_2^T$,
we uncover an order-16 nonabelian $\tilde G_{\psi} \equiv  \frac{\bD_8^{\rF,\rC \rP} \ \times \Z_4^{\rT \rF}}{\Z_2^{\rF}}$ for a 3+1d spin-1/2 Dirac field
(see the later \eq{CPT-D8-Z4-central} for explanations). 
Remarkably, the fermion parity $\Z_2^\rF$ generated by $(-1)^\rF: \psi \mapsto - \psi$ plays a crucial role
in the group extension structure 
$1 \to \Z_2^\rF \to \tilde G_{\psi}  \to G_{\phi} \to 1$. 
Thus fermionic systems reveal $\Z_2^\rF$-enriched structures richer than bosonic systems.
This means that Dirac fermion $\psi$ is in a projective rep of $G_{\phi}$, also in an (anti)linear rep of $\tilde G_{\psi}$.
(It is antilinear because $\tilde G_{\psi}$ contains the antilinear time-reversal symmetry.)

This beyond-$\Z_2$ group structure for C-P-R-T 
is mostly secretly hidden in the literature, and still not yet widely appreciated.
(However, a well-known exception is
the time-reversal symmetry can be $\Z_4^{\rT \rF} \supset \Z_2^\rF$ that $\rT^2=(-1)^\rF$ in contrast with the usual $\Z_2^\rT$ with $\rT^2=+1$,
both have applications to the classification of topological superconductors and insulators, see for instance 
\cite{RyuSPT0803.2786,  Kitaevperiod0901.2686, AIP0905.2029, Wen1111.6341, CWang1401.1142,  1406.0307CTHsiehMorimotoRyu, 
Metlitski20141406.3032, 1510.05663Metlitski, Freed1604.06527, 1711.11587GPW, HasonKomargodskiThorngren1910.14039}).
Moreover, here we stress various \emph{new nonabelian finite group structures} for the total C-P-R-T symmetries 
that have not yet been discovered previously.
Below we work through several examples in sections.

\section{3+1d spin-1/2 fermionic spinors}

First, we consider the 3+1d Dirac theory with a 4 complex component spinor field $\psi$.
We aim to carry out its C-P-R-T-$(-1) ^\rF$ structure acting on $\psi$ in detail.
It is convenient to regard the massless Dirac spinor as
two complex Weyl spinors ${\bf 2}_{L}\oplus {\bf 2}_{R}$ (left $L$ and right $R$) {rep}
in the standard Weyl basis for $\psi$ \cite{Peskin1995, Weinberg1995, Zee2003, Srednicki2007}.
Each of 4 spinor components carries different quantum numbers of
momentum ($\hat p_z$), Lorentz spin ($\hat S_z$),
and the chirality ($L$ or $R$, which is determined by helicity $\hat h= \hat p \cdot \hat S = -$ or $+$, in the massless case), shown in \Table{tab:CPRT3+1d}.

We summarize how
C, P, and T act on the spinor and its various quantum numbers intuitively in
\Table{tab:switch3+1d}:
\begin{table}[htp]
\begin{center}
\begin{tabular}{c|c|c | c| c}
\hline
 \begin{tabular}{c}
 {spinor}\\
 {component} 
\end{tabular} & $\hat p_z$ & $\hat S_z$ & $\hat h= \hat p \cdot \hat S$ &  chirality P$_{{L/R}}$\\
\hline
1st & $-$ & $+$ & $-$ & $L$ \\
2nd & $+$& $-$ & $-$ & $L$ \\
3rd & $+$ & $+$  & $+$  & $R$\\
4th & $-$ & $-$ & $+$ & $R$\\
\hline
\end{tabular}
\caption{The 4-component complex massless Dirac spinor field $\psi$ in 3+1d 
contains 8 real degrees of freedom composed from $2 \times 2 \times 2$,
chiralities (Left/Right) $\times$ $\hat S_z$ spins ($\up/\down$)
$\times$ Particle/Antiparticle. The $+$ or $-$ entry means the quantum number eigenvalue is positive or negative.}
\label{tab:CPRT3+1d}
\end{center}
\end{table} 
\begin{table}[!h]
\begin{center}
\begin{tabular}{c|c|c | c| c}
\hline
 \begin{tabular}{c}
 {discrete symmetry}\\
 {switch quantum}\\
 {numbers or not} 
\end{tabular}
 &  
 \begin{tabular}{c}
                    $p_z >0$ \\
                    $\Updownarrow$\\
                    $p_z <0$ \\
            \end{tabular}
  &  
  \begin{tabular}{c}
                    {$\hat S_z \uparrow$} \\
                    $\Updownarrow$\\
                    {$\hat S_z \downarrow$} \\
            \end{tabular}
  & 
    \begin{tabular}{c}
                    {$L$} \\
                    $\Updownarrow$\\
                    {$R$} \\
            \end{tabular}
   &   \begin{tabular}{c}
                    {\text{particle}} \\
                    $\Updownarrow$\\
                    antiparticle \\
            \end{tabular}\\
\hline
C &  &   &   & Yes \\
P / R&  Yes &  & Yes &  \\
T & Yes &  Yes  &  & \\
\hline
\end{tabular}
\caption{Agree with \eq{eq:CPT3+1d}, we show whether each spinor component and its quantum numbers are switched
under the C-P-R-T transformation.
The top horizontal row shows which quantum numbers,
and the left vertical column shows how C, P/R, or T acts. 
The ``Yes'' entry in the table means the discrete symmetry switches the quantum numbers.
The empty entry means the quantum number is preserved.
}
\label{tab:switch3+1d}
\end{center}
\end{table}\\
$\bullet$ The \emph{unitary} C switches between
the particle $\Leftrightarrow$ antiparticle,
but keeps the momentum $p_z$, the spins $\hat S_z$, and the chirality intact. 
Note that the antiparticle's 1st, 2nd, 3rd, 4th components of the 4-component spinor
have the quantum numbers of the $\hat S_z$ and chirality (opposite with respect to those of the particle's):
$(-,+,-,+)$ for $\hat S_z$, and $(R,R,L,L)$ for chirality.
See various clarifications 
in \cite{Zirnbauer2004.07107}.\\
$\bullet$ The \emph{unitary} P switches between
the momentum $p_z >0 \Leftrightarrow p_z <0$,
also switches between the chirality $L \Leftrightarrow R$, but keeps the spin $\hat S_z$ intact.\\
$\bullet$ The \emph{antiunitary} T switches between
the momentum $p_z >0 \Leftrightarrow p_z <0$
and the spin $\hat S_z$'s $\uparrow \Leftrightarrow \downarrow$, 
but keeps the chirality intact.

Below we manifest the C-P-T transformation of \Table{tab:switch3+1d} explicitly in a set of gamma matrices acting on the spinor $\psi$.
We adopt the standard Pauli matrix convention
$$
\sigma^0=
({\begin{smallmatrix}
1 &0 \\
0 &1
\end{smallmatrix}}),\;
\sigma^1=
({\begin{smallmatrix}
0 &1 \\
1 &0
\end{smallmatrix}}),\;
\sigma^2=
({\begin{smallmatrix}
0 &-\ii \\
\ii &0
\end{smallmatrix}}),\;
\sigma^3=
({\begin{smallmatrix}
1 &0 \\
0 &-1
\end{smallmatrix}}),
$$
for the gamma matrices of Clifford algebra
$\{\gamma^\mu, \gamma^\nu\}= 2 g^{\mu\nu}$ with the metric signature $(+,-,-,-)$ 
in the chiral Weyl basis:
\bea
\gamma^0 &=&
\sigma^1 \otimes \sigma^0=
{\begin{pmatrix}
0 &\sigma^0 \\
\sigma^0 &0
\end{pmatrix}}.\nn \\
\gamma^j &=&
\ii \sigma^2 \otimes \sigma^j=
{\begin{pmatrix}
0 &\sigma^j \\
-\sigma^j &0
\end{pmatrix}},
\text{ for $j=1,2,3$}.\\
\gamma^5 &=&
- \sigma^3 \otimes \sigma^0
=\ii {\gamma^0}{\gamma^1}{\gamma^2}{\gamma^3}
={\begin{pmatrix}
-\sigma^0 & 0\\
 0 & \sigma^0
\end{pmatrix}}. \nn
\eea

The \emph{active} C-P-T transformation on the fields changes $\psi$ to  $\psi'$ 
(instead of the passive transformation on coordinates),
but we adopt the primed coordinate notations, $x'_{\rP}$ and $x'_{\rT}$, introduced earlier in \eq{eq:coordinate}:
\bea
&&\rC\psi(x) 
\rC^{-1}
=\psi'_\rC(x)= - \ii \gamma^2 \psi^*(x)   
={\begin{pmatrix}
0 &{\begin{smallmatrix}
0 &-1 \\
1 &0
\end{smallmatrix}}\\
{\begin{smallmatrix}
0 &1 \\
-1 &0
\end{smallmatrix}} &0
\end{pmatrix}}\psi^*(x). 
\nn
\\
&&\rP \psi(x) \rP^{-1}
=\psi'_\rP(x)
=  \gamma^0 \psi(x'_{\rP})
={\begin{pmatrix}
0 & {\begin{smallmatrix}
1 &0 \\
0 &1
\end{smallmatrix}} 
\\
{\begin{smallmatrix}
1 &0 \\
0 &1
\end{smallmatrix}} &0
\end{pmatrix}} \psi(x'_{\rP}). 
\label{eq:CPT3+1d}\\
&&\rT\psi(x) \rT^{-1}
=\psi'_\rT(x)
=  -\gamma^1\gamma^3 \psi(x'_{\rT})
=
{\begin{pmatrix}
{\begin{smallmatrix}
0 & -1 \\
1 &0
\end{smallmatrix}} & 0 \\
0 & {\begin{smallmatrix}
0 &-1 \\
1 &0
\end{smallmatrix}}
\end{pmatrix}}\psi(x'_{\rT}). 
\nn
\\
&&(\rC \rP \rT)\psi(x) (\rC \rP \rT)^{-1}
=\psi'_{\rC \rP \rT}(x)
= \gamma^5\psi^*(-x). 
\cr
&& \rT^2=(\rC\rP)^2=(-1)^\rF. \quad \rC^2=\rP^2=(\rC \rP \rT)^2=+1. \nn
\eea
%

The
\emph{unitary} C says $\rC (z \psi(x)) \rC^{-1} =  z (- \ii \gamma^2 \psi^*(x))$
with a \emph{linear} map on a complex number $z \in \mathbb{C}$.
The C in \eq{eq:CPT3+1d} indeed agrees with \Table{tab:switch3+1d},
by taking into account that the
spin ($\hat S_z$) and chirality ($L/R$) quantum numbers of anti-particle $\psi^*$  
are opposite to that of particle $\psi$ in \Table{tab:CPRT3+1d},
namely $(-,+,-,+)$ and $(R,R,L,L)$ for each of four components of spinor $\psi^*$.
%

{The \emph{antiunitary} T actually requires a complex conjugation K to do the \emph{antilinear} map
$\rT (z \psi(x)) \rT^{-1} 
= - z^* \gamma^1\gamma^3 \psi(x'_{\rT})$. 
The complex conjugation 
K maps $z \in \mathbb{C} \mapsto \rK z \rK = z^* \in \mathbb{C}$ 
with a state-vector-basis-dependence on the Hilbert space.
But luckily these specific Weyl basis gamma matrices in \eq{eq:CPT3+1d} make this K not manifest
because all the \emph{linear} maps (i.e., $-\ii \gamma^2, \gamma^0, -\gamma^1 \gamma^3$, and $\gamma^5$)
in \eq{eq:CPT3+1d} contain only the \emph{real} coefficient matrices.
}

Clearly the Dirac spinor theory (here $d+1=3+1$) action
$
\int \dd^{d+1}x \; \bar \psi (\ii \gamma^\mu \prt_\mu - m)\psi
$,
preserves the discrete symmetry transformations in \eq{eq:CPT3+1d}.
Lo and behold, based on a chain of remarks listed below \eq{CPT-D8-Z4-central}, 
we discover the total discrete nonabelian finite group structure, of C/P/T and $(-1)^\rF$, 
summarized as $\tilde{G}_{\psi} \equiv{\frac{\bD_8^{\rF,\rC \rP} \times \mathbb{Z}_{4}^{\rT \rF} }{\Z_2^\rF}}$:\\[-6mm]
\begin{equation}
\label{CPT-D8-Z4-central}
\resizebox{76mm}{1.8mm}
{\xymatrix{
&1 \ar[d] && \\
1\ar[r]& \mathbb{Z}_{2}^{\rF} \ar[r] \ar[d]&\bD_8^{\rF,\rC \rP}  \ar[r] \ar@{^{(}->}[d]& \mathbb{Z}_{2}^{\rC} \times \mathbb{Z}_{2}^{\rP} \ar[r] &1\\
&\mathbb{Z}_{4}^{\rT \rF}  \ar[d] \ar@{^{(}->}[r]& 
\underset{\text{central product}}{\tilde{G}_{\psi} \equiv \frac{\bD_8^{\rF,\rC \rP} \times \mathbb{Z}_{4}^{\rT \rF} }{\Z_2^\rF}}
\ar@{=}[d]
& 
&\\
&\mathbb{Z}_{2}^{\rT}  \ar[d]& 
\underset{\text{central product}}{\tilde{G}_{\psi} \equiv \frac{\bQ_8^{\rF,\rC \rP,\rP\rT} \times \mathbb{Z}_{4}^{\rT \rF} }{\Z_2^\rF}}
&\\
& 1 && 
}}
\end{equation}
Let us now elaborate on \eq{CPT-D8-Z4-central} in detail step-by-step:
%
\begin{enumerate}[leftmargin=-.8mm]
\item We have $\rT^2=(-1)^\rF$ so the time-reversal $\Z_2^\rT$ and fermion parity $\Z_2^\rF$
combines to be an order-4 abelian group
$\Z_4^{\rT \rF} \supset \Z_2^\rF$ such that
the total group $\Z_4^{\rT \rF}$ sits in the group extension of
the quotient $\Z_2^\rF$ extended by the normal subgroup $\Z_2^\rF$, written as a short exact sequence:
\bea \label{eq:Z4TF}
1 \to \Z_{2}^{\rF} \to  \Z_4^{\rT \rF} \to \Z_{2}^{\rT} \to 1.
\eea

\begin{figure}[!h]
\includegraphics[width=0.9\columnwidth]{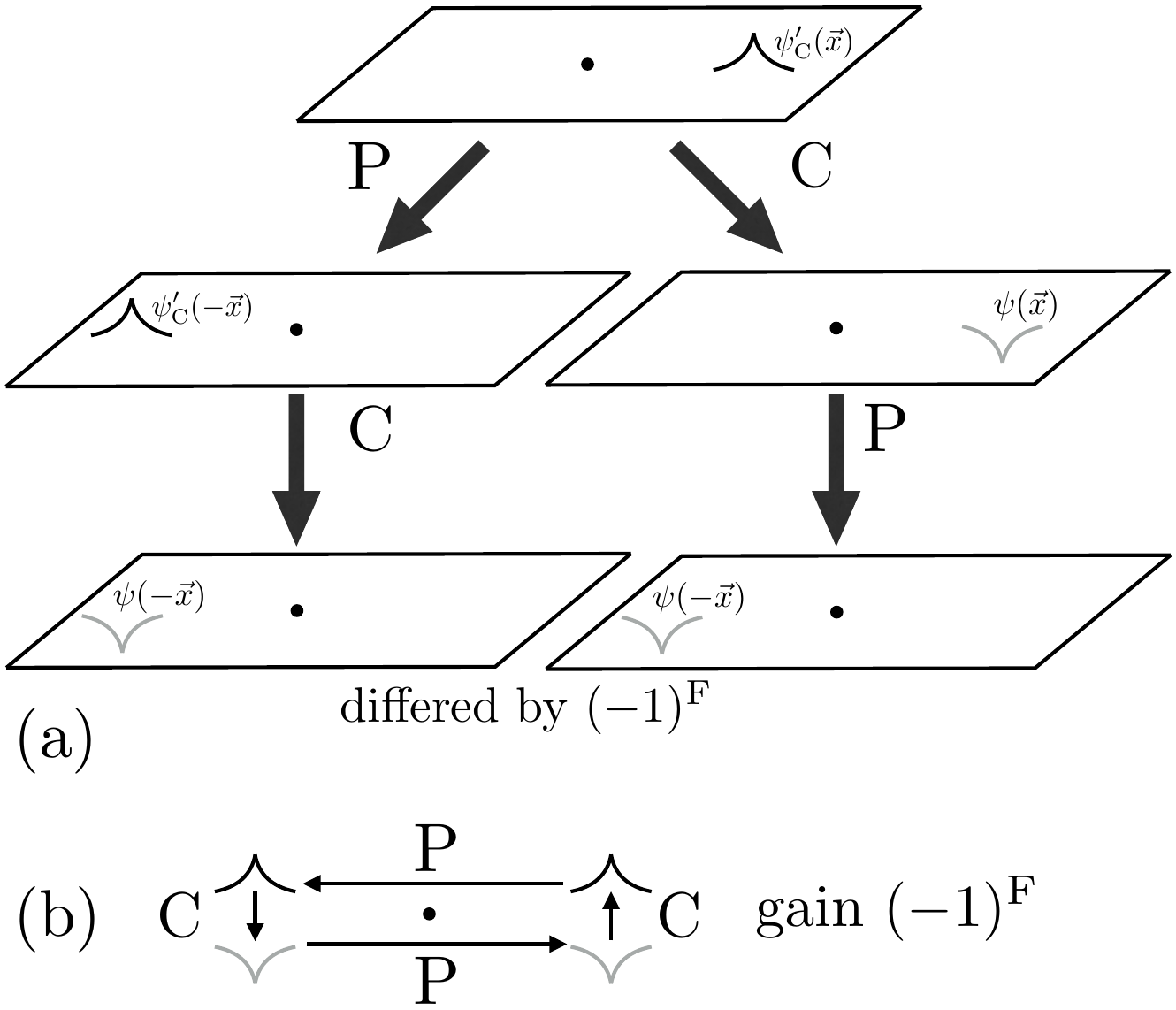}
\caption{Schematic illustrations 
(a) CP and PC act on a local Dirac fermionic excitation,
two final configurations differed by $(-1)^\rF$ due to $\rC\rP=(-1)^\rF \rP \rC$.
Namely, the following two procedures differed by a $(-1)$ sign for a Dirac fermion: 
(i) Apply P to map the particle to its mirror partner, then apply C to map the particle to its antiparticle. 
(ii) Apply C to map the particle to its antiparticle, then apply P to map the antiparticle to its mirror partner. 
More generally, the parity P here (in even spacetime dimensions) can be replaced by the reflection R.
The P or R transformation is with respect to the origin (the black dot). The white planes indicate the spatial dimensions.
The $\psi_{\rm C}'$ and $\psi$ are fermionic particle and anti-particle excitation creation operators respectively.
The convex or concave cusps represent the particle or hole excitations.
(b) A consecutive procedure $\rC\rP\rC\rP=(-1)^\rF$ gives a minus sign to a fermionic excitation.
}
\label{fig:CPCP}
\end{figure}

\item
Remarkably $\rC\rP=(-1)^\rF \rP \rC$ here, 
while
we can show $\rC \rP\psi \rP^{-1} \rC^{-1}$  $=$
$- \ii \gamma^2\gamma^0 \psi^*(x'_{\rP})$
and
$ \rP \rC\psi  \rC^{-1} \rP^{-1}$ $=$
$+ \ii \gamma^2\gamma^0 \psi^*(x'_{\rP})$ in this particular basis.
This means the C and P do not commute in the fermion parity odd  $(-1)^\rF=-1$ sector (illustrated in \Fig{fig:CPCP}),
but they commute in the bosonic  $(-1)^\rF=+1$ sector.
The $\rC$ and $\rP$ form a nonabelian finite group of order-8, a dihedral group $\bD_8$, denoted by a standard group theory notation 
via enlisting its generators (on the left) and their multiplicative properties (on the right):
\be \label{eq:D8FCP}
\mathbb{D}_8^{\rF,\rC \rP} \equiv
\<\rC \rP, \rC |(\rC \rP)^4=\rC^2=+1, \rC(\rC \rP)\rC=(\rC\rP)^{-1}\>.\quad\quad
\ee
Note that we can either understand the $\mathbb{D}_8^{\rF,\rC \rP} = \Z_4^{\rC \rP} \rtimes \Z_2^\rC$
via the group extension
$
1 \to \Z_4^{\rC \rP}  \to \mathbb{D}_8^{\rF,\rC \rP} \to \Z_2^\rC \to 1
$ with the order-4 $ \Z_4^{\rC \rP}$ sits at the normal subgroup and the $\Z_2^\rC$ (or $\Z_2^\rP$) sits at the quotient;
or we can understand the $\bD_8^{\rF,\rC \rP}$ as
the quotient $ \mathbb{Z}_{2}^{\rC} \times \mathbb{Z}_{2}^{\rP}$ extended by the fermion parity $\Z_2^\rF$ as
another group extension:
\bea \label{eq:D8FCP}
1\to  \Z_2^\rF \to \bD_8^{\rF,\rC \rP}  \to  \mathbb{Z}_{2}^{\rC} \times \mathbb{Z}_{2}^{\rP} \to 1. 
\eea
Note that $({\rC \rP})^2=\rT^2=(-1)^{\rF}$.

\item The \eq{CPT-D8-Z4-central}'s
vertical and horizontal group extensions are already explained in \eq{eq:Z4TF} and \eq{eq:D8FCP}
as two short exact sequences.
The standard notation of the inclusion $``\hookrightarrow''$ in
$G_{\text{sub}} \hookrightarrow G$ means that $G$ contains $G_{\text{sub}}$ as a subgroup.
This order-16 nonabelian finite group
$\tilde{G}_{\psi} \equiv {\frac{\bD_8^{\rF,\rC \rP} \times \mathbb{Z}_{4}^{\rT \rF} }{\Z_2^\rF}}$
 contains both $\bD_8^{\rF,\rC \rP}$
and $\mathbb{Z}_{4}^{\rT \rF}$ subgroups, as their inclusion notations ($\hookrightarrow$) suggest.
The 
$\tilde{G}_{\psi}$ is the \emph{central product} between
${\bD_8^{\rF,\rC \rP} \times \mathbb{Z}_{4}^{\rT \rF} }$ mod out their common ${\Z_2^\rF}$ center subgroup, as their 
${\Z_2^\rF}$ is identical. 
Amusingly this $\tilde{G}_{\psi}$ is isomorphic to the 16-element rank-2 matrix group known as Pauli group
$\equiv \<\sigma^1,\sigma^2, \sigma^3\>$
generated by Pauli matrices that act on the 2-dimensional Hilbert space of 1 qubit.

\item Now we show 
$\tilde{G}_{\psi} \equiv {\frac{\bD_8^{\rF,\rC \rP} \times \mathbb{Z}_{4}^{\rT \rF} }{\Z_2^\rF}}
={\frac{\bQ_8^{\rF,\rC \rP,\rP\rT} \times \mathbb{Z}_{4}^{\rT \rF} }{\Z_2^\rF}}$ group isomorphism,
which basically says two facts:
(1) the first group generated by C, P, T, and the second group 
generated by CP, PT, CT, and T, are exactly the same order-16 nonabelian group,
(2) an order-8 quaternion group 
\be \hspace{-3mm}
\mathbb{Q}_8^{\rF,\rC \rP,\rP\rT} 
=\<\rC \rP, \rP \rT, \rC \rT |
(\rC \rP)^2=( \rP \rT)^2=(\rC \rT)^2=(-1)^\rF\>
\ee is
generated by $\mathbf{i} = \rC \rP, \mathbf{j}=  \rP \rT$, and $\mathbf{k} = \rC \rT$ via a standard notation
$\bQ_8
=
\< \mathbf{i}, \mathbf{j}, \mathbf{k} | \mathbf{i}^2=\mathbf{j}^2=\mathbf{k}^2=\mathbf{ijk} =-1
\>$. 
\item Because the Dirac spinor $\psi$ sits in the complex ${\bf 2}_{L}\oplus {\bf 2}_{R}$ {rep} 
of spacetime symmetry Spin(3,1), we can ask: How does the order-16 nonabelian finite group
fit into the Dirac theory's spacetime-internal symmetry group 
\bea \label{eq:spacetime-internal-symmetry-group}
G_{\text{spacetime} \atop \text{-internal}} ={G_{\text{spacetime} }  \ltimes_{N} G_\text{internal}}
\eea
(the semi-direct product mod out the common normal subgroup $N$ is denoted as ``$\ltimes_{N}$'')? 
In Minkowski signature flat spacetime, 
we have
$G_{\text{spacetime} }=\Pin(d,1)$, 
which not only is a double-cover of $\rO(d,1)$, but
 also contains a normal subgroup $\Spin(d,1)$. All these
$\Pin(d,1)$, $\rO(d,1)$, and $\Spin(d,1)$
sit inside the group extension:\\[-6mm]
\begin{equation}
\label{eq:group-extension}
\resizebox{56mm}{1.5mm}{
\xymatrix{
&1 \ar[d] &1 \ar[d] &  \\
& \mathbb{Z}_{2}^{\rF} \ar[d] & \mathbb{Z}_{2}^{\rF} \ar[d]  & \\
1\ar[r]& \Spin(d,1)\ar[r] \ar[d]&\Pin^{}(d,1) \ar[d] \ar[r] & \mathbb{Z}_{2}^{T}  \ar[r]  &1\\
1\ar[r]& \SO(d,1)_{} \ar[r] \ar[d]&\rO(d,1)\ar[r]^{{\det}=\pm 1}  \ar[d]& \mathbb{Z}_{2}^{T} \ar[r]  &1\\
& 1 &1 & 
}
}
\end{equation}
Note that a special orthogonal 
$\SO(d,1)$ contains two components (
$\pi_0(\SO(d,1))=\Z_2$), the proper orthochronous 
Lorentz group  
$\SO^+(d,1)$ and another component
that can be switched via the simultaneous R and T (say $\Z_2^{\rR \rT}$).  
Thus,
\bea
1 \to \SO^+(d,1) \to &\SO(d,1)& \to \Z_2^{\rR\rT} \to 1,\cr
1 \to \SO^+(d,1) \to &\rO(d,1)& \to \Z_2^{\rR} \times \Z_2^{\rT} \to 1.
\eea
Note that here we choose the $\Pin^{}(d,1)$ instead of $\Pin^{}(1,d)$ because
a generic non-isomorphism
$\Pin^{}(d,1) \not \cong \Pin^{}(1,d)$, while the former has
their $\rT^2$ and Clifford algebra as  \cite{Freed1604.06527} 
$$
\rT^2=(-1)^\rF, \; \text{Cliff}_{d,1}: e_0^2=-1,\; e_j^2=1, \text{ with } j=1,\dots,d,
$$ 
the later has a different property, not we required here:
$$
\rT^2=+1, \quad \text{Cliff}_{1,d}: e_0^2=1,\; e_j^2=-1, \text{ with } j=1,\dots,d.
$$ 
In short, $\Pin^{}(d,1)$ not only contains the $\Z_2^{\rF}$ center,
but also contains four connected components,
i.e., $\pi_0(\Pin^{}(d,1) )=\Z_2 \times \Z_2$, same as $\pi_0(\rO^{}(d,1) )=\Z_2 \times \Z_2$,
disconnected from each other flipped by $\Z_2^{\rR}$ and $\Z_2^{\rT}$.
\item The three discrete subgroups, $\Z_2^{\rR}$, $ \Z_{2}^{\rT},$ and $ \Z_{2}^{\rF}$ 
are found as some normal subgroup or quotient group in \eq{eq:group-extension}.
But where is the missing charge conjugation $\Z_{2}^{\rC}$?

$\bullet$ In general, the charge conjugation is better defined mathematically \cite{Freed1604.06527} as 
a new element of the
extended group in the CRT theorem, 
acting by \emph{conjugate linear} (\emph{antilinear}) maps on the Hilbert
space of statevectors. 
This follows Wigner's theorem on symmetries of a quantum system \cite{wigner2012group}: 
any transformation of projective Hilbert space that preserves the absolute value of the inner products 
can be represented by a \emph{linear} or \emph{antilinear} transformation of Hilbert space, which is unique up to a phase factor.


$\bullet$ In a particular narrow-minded 
purpose here, we can include naturally the internal symmetry $G_{\text{internal}} = \U(1)$
into the full spacetime-internal symmetry of Dirac theory's
$G_{\text{spacetime} \atop \text{-internal}}=\Pin(d,1) \ltimes_{\Z_2^\rF} \U(1)$ in \eq{eq:spacetime-internal-symmetry-group}
such that the charge conjugation C is the complex conjugation of the U(1),
which maps $g = \e^{\ii q \theta} \in \U(1)$
to $g^* = \e^{-\ii q \theta} \in \U(1)$.
Thus the charge conjugation generates the outer automorphism of the U(1):
${\rm Out}(\U(1))= \Z_{2}^{\rC}$.

In 3+1d, the outer automorphism of $G_{\text{spacetime} \atop \text{-internal}}$ still is: 
${\rm Out}(\Pin(3,1) \ltimes_{\Z_2^\rF} \U(1))=\Z_{2}^{},$
the only natural charge conjugation available.
%

The benefit of this viewpoint is that $G_{\text{spacetime} \atop \text{-internal}}=\Pin(d,1) \ltimes_{\Z_2^\rF} \U(1)$ relates 
to the so-called AII class topological insulator's symmetry group
in the Wigner-Dyson-Altland-Zirnbauer symmetry classification \cite{wigner1951, Dyson1962, AltlandZirnbauer9602137}.

$\bullet$ In summary of the above, we put four 
$\Z_2$ groups together:
$\Z_2^{\rP}$, $\Z_2^{\rR}$, $\Z_2^{\rT}$
into disconnected components of \eq{eq:group-extension},
and the $\Z_2^{\rC}$ can be introduced either  
(1) generally by a conjugate linear map on the Hilbert
space of statevectors, or
(2) narrowly by an outer automorphism of $G_{\text{internal}}$
or $G_{\text{spacetime} \atop \text{-internal}}$. 
Then, the order-16 group
can be fitted into both \eq{CPT-D8-Z4-central}
and \eq{eq:group-extension}'s framework.

$\bullet$ We can also view the $\tilde{G}_{\psi} \equiv {\frac{\bD_8^{\rF,\rC \rP} \times \mathbb{Z}_{4}^{\rT \rF} }{\Z_2^\rF}}$ 
extended from the bosonic $G_{\phi}\equiv\Z_2^{\rC} \times \Z_2^{\rP} \times \Z_2^{\rT}$ via a $\Z_2^{\rF}$ extension: 
\bea \label{eq:projective-linear-representation}
1 \to \Z_2^{\rF} \to {\frac{\bD_8^{\rF,\rC \rP} \times \mathbb{Z}_{4}^{\rT \rF} }{\Z_2^\rF}} \to \Z_2^{\rC} \times \Z_2^{\rP} \times \Z_2^{\rT} \to 1.
\eea
Then the spin-0 boson $\phi$ sits at an {(\emph{anti})\emph{linear representation}} of $G_{\phi}$,
but the spin-1/2 Dirac fermion $\psi$ sits at a \emph{projective representation} of $G_{\phi}$. 
The $\psi$ carries fractional quantum numbers of $G_{\phi}$ is in fact in an 
{(\emph{anti})\emph{linear representation}} of $\tilde{G}_{\psi}$.
The spinor $\psi$ is thus a fractionalization of a scalar $\phi$.
{The \emph{symmetry extension} \cite{Wang2017locWWW1705.06728}
as $1 \to \Z_2^{\rF} \to \tilde{G}_{\psi} \to G_{\phi} \to 1 $
implies that whether $\psi$ may or may not have 't Hooft anomaly in $G_{\phi}$, but  
$\psi$ can become anomaly-free via the pullback to $\tilde{G}_{\psi}$.}
\hfill $\blacksquare$

\item
In addition, we can study other similar spacetime-internal symmetry, compatible
with $G_{\text{spacetime}}$ contains Lorentz (boost and rotation) symmetry
and $G_{\text{internal}} = \U(1)$
while they both share $\Z_2^{\rF}$. This can be done, by solving the group extension  \cite{Freed1604.06527, WWZ1912.13504}:
$1 \to \rO(d,1) \to  G_{\text{spacetime} \atop \text{-internal}} \to \U(1) \to 1,$
and enumerating the solutions of $G_{\text{spacetime} \atop \text{-internal}}$,
based on Minkowski or Euclidean notations:
\bea 
\Pin(d,1) \ltimes_{\Z_2^\rF} \U(1) \text{ or }  \Pin^{\tilde c +} \equiv \Pin^+ \ltimes_{\Z_2^\rF} \U(1) &:&\text{AII},\cr
\Pin(1,d) \ltimes_{\Z_2^\rF} \U(1) \text{ or }  \Pin^{\tilde c -} \equiv \Pin^- \ltimes_{\Z_2^\rF} \U(1) &:& \text{AI}, \label{eq:Pin-A-sym-class}\quad\quad\;\\
\Pin(d,1) \times_{\Z_2^\rF} \U(1) \text{ or } \Pin^c \equiv \Pin^{\pm} \times_{\Z_2^\rF} \U(1) &:& \text{AIII}.\nn
\eea 
These groups are known to be compatible with 
AII, AI, and AIII symmetry classifications of quantum (e.g., condensed or nuclear) matters \cite{wigner1951, Dyson1962, AltlandZirnbauer9602137}.
The AI and AII have $\rT^2=+1$ and $\rT^2=(-1)^\rF$ respectively, 
the anti-unitary T does \emph{not} commute with a charge-like (operator $\hat q$) U(1):
\bea
\rT U_{\U(1)}=  U_{\U(1)}^{-1} \rT, \quad  \text{ namely }  \rT \e^{\ii \hat q \theta} = \e^{-\ii \hat q \theta} \rT,
\eea
known also as the symmetry of topological insulators.\\
For AIII, regardless $\rT^2=+1$ or $(-1)^\rF$,
the anti-unitary T commutes with an isospin-like (operator $\hat s$) U(1):
\bea \label{eq:U1s}
\rT U_{\U(1)}=  U_{\U(1)} \rT, \quad \text{ namely } \rT \e^{\ii \hat s \theta} = \e^{\ii \hat s \theta} \rT,
\eea
known also as the symmetry of topological superconductors.
Note that $\rT \ii \rT^{-1}= -\ii$, $\rT \hat q \rT^{-1}=\hat q$, and $\rT \hat s \rT^{-1}=-\hat s$.\\
%
%
$\bullet$ The AII case has a total 
$\tilde{G}_{\psi}={\frac{\bD_8^{\rF,\rC \rP} \times \mathbb{Z}_{4}^{\rT \rF} }{\Z_2^\rF}}$ in \Eq{CPT-D8-Z4-central}.

$\bullet$ The AI case has $\rT^2=+1$, 
so we replace \eq{CPT-D8-Z4-central}'s  $\Z_4^{\rT \rF}$ by another subgroup $\Z_2^\rF \times \Z_2^\rT$ instead.
Then \eq{CPT-D8-Z4-central} reduces to a different order-16
nonabelian $\tilde{G}_{\psi}={\bD_8^{\rF,\rC \rP} \times \mathbb{Z}_{2}^{\rT} }$.

$\bullet$ The AIII case has a subtle U(1) and T relation given by \eq{eq:U1s}, e.g.,
one can realize this new $\rT'$ as the combined $\rT'=\rC\rT$ \cite{1510.05663Metlitski, 1711.11587GPW} of \eq{eq:CPT3+1d}.
We leave this and other symmetry realizations in upcoming works \cite{to-appear}.
\hfill $\blacksquare$ 

\item {\bf Majorana fermion}:
Other than the Dirac spinor $\psi$ discussed above,
we can ask what happens to Majorana spinor? Once we impose the Majorana condition
$$
\rC\psi(x) 
\rC^{-1} = \psi_{\rC}(x) = - \ii \gamma^2 \psi^*(x)
=  \psi (x),
$$
the $\Z_2^{\rC}$ acts trivially as an identity on Majorana spinor. Therefore,
we shall reduce the total group structure to P-R-T-$(-1)^{\rF}$ without C.
Then \eq{CPT-D8-Z4-central}'s total group $\tilde{G}_{\psi}$ reduces to an order-8
abelian group, ${\Z_2^{\rP} \times \mathbb{Z}_{4}^{\rT \rF} }$ for the AII case,
and  $\Z_2^{\rF}  \times \Z_2^{\rP}  \times \Z_2^{\rT}$ for the AI case.
\hfill $\blacksquare$ 
\end{enumerate}

\section{1+1d spin-1/2 fermionic spinors}

Now we move on to the C-P-R-T fractionalization structure for 1+1d relativistic fermions.\\
\noindent
{\bf Dirac fermion}:
We can regard a 1+1d massless Dirac spinor $\psi$ as
two complex Weyl spinors in ${\bf 1}_L\oplus {\bf 1}_R$ (left L + right R) {rep},
easily seen in the Weyl basis
gamma matrices: 
$$
\gamma^0=\sigma^1=
({\begin{smallmatrix}
0 &1 \\
1 & 0
\end{smallmatrix}}),\;
\gamma^1=\ii\sigma^2=
({\begin{smallmatrix}
0 &1 \\
-1 &0
\end{smallmatrix}}),\;
\gamma^5=\gamma^0\gamma^1
=
({\begin{smallmatrix}
-1 &0 \\
0 &1
\end{smallmatrix}}).
$$
The \emph{active} C-P-T transformation on $\psi$ gives:
\bea
&&\rC\psi(x) 
\rC^{-1}=\psi'_\rC(x)=  \gamma^5 \psi^*(x)   
=({\begin{smallmatrix}
-1 &0 \\
0 &1
\end{smallmatrix}})\psi^*(x). 
\nn
\\
&&\rP \psi(x) \rP^{-1} =\psi'_\rP(x)=  \gamma^0 \psi(x'_{\rP})
=({\begin{smallmatrix}
0 &1 \\
1 & 0
\end{smallmatrix}}) \psi(x'_{\rP}). 
\\
&&\rT\psi(x) \rT^{-1} =\psi'_\rT(x)=  \gamma^0  \psi(x'_{\rT})
=
({\begin{smallmatrix}
0 &1 \\
1 & 0
\end{smallmatrix}})  \psi(x'_{\rT}). 
\nn
\\
&&(\rC \rP \rT)\psi(x) (\rC \rP \rT)^{-1} =\psi'_{\rC \rP \rT}(x) =
\gamma^5\psi^*(-x). 
\cr
&&\rC^2=\rP^2= \rT^2=(\rC \rP \rT)^2=+1. \quad (\rC\rP)^2=(-1)^\rF.\nn
\eea
$\bullet$ Remarkably $\rC\rP=(-1)^\rF \rP \rC$, 
so we still have \eq{eq:D8FCP}'s $\mathbb{D}_8^{\rF,\rC \rP}$.\\
$\bullet$ Again T is anti-unitary, so precisely 
$\rT (z \psi(x)) \rT^{-1} 
=z^* \gamma^0  \psi(x'_{\rT})$, 
but luckily the complex conjugation K is not manifest in this gamma matrix basis.
Since $\rT^2=+1$, the $\Z_4^{\rT\rF}$ in
\eq{CPT-D8-Z4-central} is replaced by the $\Z_2^{\rF} \times \Z_2^{\rT}$.
\\ 
$\bullet$ PT commutes with every group element,
so we derive that the order-16 total group is
$\mathbb{D}_8^{\rF,\rC \rP} \times \Z_2^{\rP \rT}$.
This particular case is within AI case in \eq{eq:Pin-A-sym-class},
we leave other spacetime-internal symmetry realizations (e.g., AII, AIII) in upcoming works \cite{to-appear}.
\hfill $\blacksquare$\\

\noindent
{\bf Majorana fermion}:
A 1+1d Majorana spinor
imposes the condition
$$
\rC\psi(x) 
\rC^{-1} = \psi_{\rC}(x) =  \gamma^5 \psi^*(x)
=  \psi (x),
$$
the $\Z_2^{\rC}$ acts trivially as an identity on the \emph{real} Majorana spinor.
Then we reduce the \eq{CPT-D8-Z4-central}'s total group to
an order-8 group
 $\Z_2^{\rF}  \times \Z_2^{\rP}  \times \Z_2^{\rT}$.
 \hfill $\blacksquare$


\section{0+1d Majorana fermion zero modes}

Kitaev's fermionic chain \cite{Kitaev2001chain0010440} is a 1+1d nonrelativistic quantum system,
hosting a Majorana zero mode on each open end of 0+1d boundary.
The 0+1d low energy effective boundary action is $\int \dd t \chi \ii \partial_t \chi$ for each 0+1d real Majorana fermion $\chi$.
There is no parity P in 0+1d, and no C for the real Majorana.
When the bulk of $k$ fermionic chains with $k \mod 8 \neq 0$
are protected by $G=\Z_2^\rF \times \Z_2^\rT$ symmetry,
the $k$-boundary's zero modes 
are not gappable (with the dimension of Hilbert space as $2^{\frac{k}{2}}$) 
as long as $G$ is preserved due to the 't Hooft anomaly in $G$ is classified by $k \in \Z_8$ \cite{FidkowskifSPT1,FidkowskifSPT2}.
\Refe{Gu1308.2488, Behrends1908.00995, MonteroVafa2008.11729,
PrakashJW2011.12320, PrakashJW2011.13921, TurzilloYou2012.04621, DelmastroGaiottoGomis2101.02218} suggest that at $k=2$ (or $k = 2 \mod 4$ in general) has 
various supersymmetric quantum mechanical interpretations. 
Concretely, we follow \Refe{PrakashJW2011.13921}, 
which shows this boundary can realize an extended symmetry
$\tilde G = \bD_8^{\rF,\rT}= \Z_4^\rT \rtimes \Z_2^\rF$. 
The 2-dimensional Hilbert space 
$\cH =\{ | \rB \rangle, | \rF \rangle\}  
= \cH_\rB \oplus \cH_\rF$
has a bosonic and a fermionic ground state, 
say $| \rB \rangle = \big(\begin{smallmatrix}1\\ 0 \end{smallmatrix}\big)$
and $| \rF \rangle = \big(\begin{smallmatrix} 0 \\ 1 \end{smallmatrix}\big)$.
The fermion parity $(-1)^{\rF}=  \big(\begin{smallmatrix} 1 & 0\\ 0 & -1  \end{smallmatrix}\big)= \sigma^3$
and the time-reversal $\rT = \big(\begin{smallmatrix} 0 & -\ii\\ \ii & 0  \end{smallmatrix}\big) \rK = \sigma^2 \rK$
do not commute, i.e., $(-1)^{\rF} \rT (-1)^{\rF} = \rT^{-1}= - \rT$. Also $\rT^2 = - \sigma^0 = -1$ and $\rT^4 = +1$.
This example can be interpreted as  a generalization of \emph{symmetry extension} \cite{Wang2017locWWW1705.06728} (in contrast to \emph{symmetry breaking})
to cancel (or trivialize) the $k=2$ anomaly in $G$ by a \emph{supersymmetry extension} pullback to $\tilde G$ \cite{PrakashJW2011.13921}.
 \emph{Supersymmetry extension} means that there exists some symmetry generator (here T) such that this generator switches 
between bosonic $| \rB \rangle$ and fermionic  $| \rF \rangle$ sectors,
thus this generator does not commute with the fermion parity $(-1)^{\rF}$.
It can be also understood as
a T-fractionalization from an order-4 abelian $G=\Z_2^\rF \times \Z_2^\rT$ (with $\rT^2= +1$)  to
an order-8 nonabelian $\tilde G=\bD_8^{\rF,\rT}= \Z_4^\rT \rtimes \Z_2^\rF$ (with $\rT^2= -1$ and $\rT^4= +1$).
\hfill $\blacksquare$

If we change the bulk symmetry to be protected by a $G=\Z_4^{\rT \rF}$,
then \Refe{PrakashJW2011.13921} finds that the $k=2$ Majorana zero mode anomaly
can be canceled  (or trivialized) by a \emph{supersymmetry extension} pullback to an order-16 nonabelian group $\tilde G =\bM_{16}$ \cite{PrakashJW2011.13921}.
It can be also understood as
a T-fractionalization from an order-4 abelian $G$ (with $\rT^2= (-1)^{\rF}$ and $\rT^4= +1$)  to
$\bM_{16}$
(with $\rT^4= -1$ and $\rT^8= +1$ \cite{Gu1308.2488, PrakashJW2011.13921}).
\hfill $\blacksquare$

\section{3+1d spin-1 Maxwell or Yang-Mills gauge theory}

We briefly analyze C-P-R-T group structure for the spin-1 gauge theories, pure U(1) Maxwell or SU(N) Yang-Mills (YM) theories  
of 3+1d actions $\int \Tr(F \wedge \star F)
- \frac{\theta}{8 \pi^2} {g}^2
 \Tr( F \wedge F )$ of a 2-form field strength $F = \dd a - \ii { g} a \wedge a$ with a $\theta$-term.
We will see that generalized global symmetries \cite{Gaiotto2014kfa1412.5148}
(i.e., 1-form symmetries $G_{[1]}$ that act on 1d Wilson or 't Hooft line operators
in contrast to 0d point particle operators) 
can enrich the group structure.
Follow the notations of \cite{WanWWZ1904.00994}, the \emph{active} C-P-T transformations act 
on spin-1 gauge bosons in terms of 1-form gauge field,  
$a = a_\mu \dd x^\mu  = a_0 \dd t + a_j \dd x^j =(a_0^{\al} \dd t + a_j^{\al} \dd x^j)T^{\al}$
with the real-valued four-vector component (namely $a_\mu^{\al} \in \R$) and the
hermitian Lie algebra generator (namely the hermitian conjugate $T^{\al \dagger}=T^{\al}$ and a real Lie structure constant $f^{\al \bt \gamma} \in \R$
in the commutator $[T^{\al},T^{\bt}]=\ii f^{\al \bt \gamma} T^{\gamma}$),  
as:
\bea \label{eq:CPT-spin-1}
&& \hspace{-6mm}\rC a_\mu^{\al}(x)
\rC^{-1}=
{\mp}
(a_0^{\al}(x), a_j^{\al}(x) ), \;   {\rC T^{\al}  \rC^{-1}=  T^{{\al}}}. \cr
&& \hspace{-6mm} \rP a_\mu^{\al}(x) 
\rP^{-1}=(a_0^{\al}(x'_{\rP}), - a_j^{\al}(x'_{\rP}) ), \;   \rP T^{\al}  \rP^{-1}=  T^{{\al}} . \cr
&& \hspace{-6mm} \rT a_\mu^{\al}(x)
\rT^{-1}=({\pm} a_0^{\al}(x'_{\rT}),  {\mp}  a_j^{\al}(x'_{\rT}) ), \;   {\rT T^{\al}  \rT^{-1}=  T^{{\al}*}}. \cr
&& \hspace{-6mm} 
\rC\rT a_\mu^{\al}(x)
(\rC\rT)^{-1}=({-} a_0^{\al}(x'_{\rT}),  {+} a_j^{\al}(x'_{\rT}) ). 
\cr
&& \hspace{-6mm} 
\rC\rP\rT a_\mu^{\al}(x) 
(\rC\rP\rT)^{-1}={-} 
(a_0^{\al}(-x),  a_j^{\al}(-x) ). 
\eea
{The gauge field associated with a real symmetric Lie algebra (namely the complex conjugate ${T^{\alpha *}}={T^{\alpha}}$) has the upper version of the sign choices.
The gauge field associated with an imaginary antisymmetric Lie algebra (namely ${T^{\alpha *}}=-{T^{\alpha}}$) has the lower version of the sign choices.}
However, overall, we can \emph{rewrite} the C-P-T symmetries on the combined $a_\mu= a_\mu^{\al}T^\al$ from \eq{eq:CPT-spin-1} equivalently as:
\bea
&& \hspace{-6mm}\rC a_\mu(x)
\rC^{-1}
=(a_0^{\al}(x), a_j^{\al}(x) )(-T^{{\al*}})
=-a^*_\mu(x). \cr
&& \hspace{-6mm}\rP a_\mu(x)
\rP^{-1}
= (a_0^{}(x'_{\rP}), - a_j^{}(x'_{\rP}) ).
\cr
&& \hspace{-6mm}\rT a_\mu(x)
\rT^{-1} = (a_0^{}(x'_{\rT}), - a_j^{}(x'_{\rT}) ). \cr
&& \hspace{-6mm}\rC\rT a_\mu(x)
(\rC\rT)^{-1} =({-} a_0^{*}(x'_{\rT}),  {+} a_j^{*}(x'_{\rT}) ).\cr
&& \hspace{-6mm} 
\rC\rP\rT a_\mu^{}(x) 
(\rC\rP\rT)^{-1}={-} 
(a_0^{*}(-x),  a_j^{*}(-x) )
= -a_\mu^{*}(-x) . \quad
\eea

Other than C-P-R-T symmetries (manifest at $\theta= 0, \pi$),
 the pure U(1) gauge theory has 1-form electric and magnetic symmetries, denoted as $\U(1)_{[1]}^{e} \times \U(1)_{[1]}^{m}$,
while the pure SU(2) YM has a 1-form electric symmetry $\Z_{2,[1]}^{e}$  \cite{Gaiotto2014kfa1412.5148}.
It can be shown that \emph{kinematically},
the U(1) gauge theory has 
$$
(\U(1)_{[1]}^{e} \times \U(1)_{[1]}^{m}) \rtimes \Z_2^{\rC}
$$
and 
where $\Z_2^{\rP} \times \Z_2^{\rT}$ are contained in the Lie group $\rO(d,1)$;
the SU(2) YM has instead
$
\Z_2^{\rP} \times \Z_2^{\rT} \times \Z_{2,[1]}^{e} \subset  \rO(d,1) \times \Z_{2,[1]}^{e}
$
(no $\Z_2^{\rC}$ due to no SU(2) outer automorphism) which fermionic/bosonic extension is studied carefully in \cite{WanWWZ1904.00994} also in \cite{WWZ1912.13504}.
These global symmetries C-P-R-T-$G_{[1]}$ are preserved \emph{kinematically} at $\theta= 0$ and $\pi$, 
but the gauge \emph{dynamical} fates (spontaneously symmetry breaking or not) are highly constrained by their 't Hooft anomalies of higher symmetries.
(These 't Hooft anomalies are firstly discovered in \cite{Gaiotto2014kfa1412.5148, Gaiotto2017yupZoharTTT1703.00501}, 
 later found to be captured by
precise invertible topological QFTs via cobordism invariants by \cite{Wan2018zql1812.11968, WanWWZ1904.00994}. 
Dynamical constraints of these anomalies are explored in particular by \cite{WanWWZ1904.00994, CordovaCO2019}.)

We leave additional analysis and other general gauge groups of gauge theories 
(see examples in \Refe{AitkenChermanUnsal1804.05845} for SU($N$) YM with $N >2$, and Ref.~\cite{CordovaHsinSeiberg1712.08639, HsinShao1909.07383} for 2+1d) for future works \cite{to-appear}.

\section{Applications} 

As applications, we briefly apply the above results to physical pertinent systems.
\begin{enumerate}[leftmargin=-.8mm]
\item For any proposed duality between two seemingly different QFTs, 
their global symmetries must be matched. So the C-P-T fractionalization provides
a constraint to verify the duality. 
\item Quantum electro-/chromo-dynamics
 (QED$_4$/QCD$_4$): 
 
$\bullet$ For Dirac fermions coupled to $\U(1)$ background fields (which $\U(1) \supset \Z_2^{\rF}$, 
the full spacetime-internal symmetry contains $\Pin^{\tilde c +}$ in \eq{eq:Pin-A-sym-class} and  
$\tilde{G}_{\psi}={\frac{\bD_8^{\rF,\rC \rP} \times \mathbb{Z}_{4}^{\rT \rF} }{\Z_2^\rF}}$).
By dynamically gauging the $\U(1)$, 
 the outcome QED$_4$ reduces the $\Pin^{\tilde c +}$ to $\rO(3,1)$ 
 while reduces the $\tilde{G}_{\psi}$ to $\Z_2^{\rC} \times \Z_2^{\rP} \times \Z_2^{\rT}$.
However, if the Dirac fermion has a large mass at ultraviolet (UV), 
at infrared (IR) there could be new emergent 1-form symmetries \cite{Gaiotto2014kfa1412.5148} 
(whose charged objects are 1-dimensional Wilson or 't Hooft lines)
which do not commute with the $\Z_2^{\rC}$.

$\bullet$ Dirac fermions can be in the fundamental or adjoint {reps} of $\SU(2)$
when coupling to $\SU(2)$ gauge fields.
In the case of the fundamental {rep}, $\SU(2) \supset \Z_2^{\rF}$, 
so the fundamental QCD$_4$ obtained by gauging $\SU(2)$ reduces 
$\tilde{G}_{\psi}$ to $\Z_2^{\rC} \times \Z_2^{\rP} \times \Z_2^{\rT}$.
However, for the adjoint {rep}, $\SU(2) \not\supset \Z_2^{\rF}$,
the resulting adjoint QCD$_4$ keeps the same order-16 $\tilde{G}_{\psi}$.
In fact, this C-P-T fractionalization $\tilde{G}_{\psi}$ can provide a constraint
to verify the UV-IR duality between the UV adjoint QCD$_4$ theory 
and the IR Dirac fermion theory 
previously studied in \cite{Anber2018iof1805.12290, Cordova2018acb1806.09592DumitrescuClay, BiSenthil1808.07465, Wan2018djlW2.1812.11955}.

$\bullet$ For Dirac fermions coupled to $\SU(3)$  in the fundamental rep
(which $\SU(3) \not\supset \Z_2^{\rF}$),
the resulting real-world $\SU(3)$ QCD$_4$
indeed can keep this C-P-T fractionalization order-16 $\tilde{G}_{\psi}$.  
Moreover, the CPT theorem and Vafa-Witten theorem \cite{VafaWittenPRL1984xg} say that CPT and P 
cannot be spontaneously broken in a vector-like QCD theory.
If the strong CP problem further indicates that the CP (thus T) is not violated in the real-world QCD$_4$
(namely, say $\theta=0$ for the $\theta$-term $\frac{\theta}{8 \pi^2} {g}^2 \Tr( F \wedge F )$),
then all discrete C-P-T are preserved which implies 
that the order-16 $\tilde{G}_{\psi}$ can be preserved in the vacuum of the real-world QCD$_4$,
at least within the strong force sector.

Of course, the weak force sector breaks P and CP, so $\tilde{G}_{\psi}$ is still violated within the full Standard Model.
\end{enumerate}

\section{Fractional Spin-Statistics and CPT}

Since the early studies by Pauli \cite{Pauli1940}, and by
Schwinger-Pauli-L\"uder \cite{SchwingerPhysRev.82.914, Pauli1955, Pauli1957, Luders1954, LUDERS19571, StreaterWightman1989}, 
physicists are intrigued by the subtle relation between the spin-statistics theorem and the CPT theorem.
Some observations and comments are in order:

\noindent
$\bullet$ We were well-informed that quantum excitations in 2+1d, called anyons, 
can have the fractional spin $s$ (self-statistics gives a Berry phase $\e^{\ii 2 \pi s}$) and also abelian or nonabelian statistics (mutual statistics), see the reviews \cite{Wilczek1990BookFractionalstatisticsanyonsuperconductivity, Nayak0707.1889}.\\
\noindent
$\bullet$
In higher dimensions (3+1d or above), there are
no 0d particle-like anyons (of 1d worldline) with fractional statistics;
but there are extended objects (1d loop-like anyonic strings on 2d worldsheets, or $n$d branes on $(n+1)$d worldvolumes) 
that can also have fractional statistics, either abelian or nonabelian statistics \cite{WangLevin1403.7437,JiangMesarosRan1404.1062, WW1404.7854}---
when those world-trajectories of these objects forming nontrivial mathematical \emph{link invariants} in the spacetime \cite{WangWenYau1602.05951,Putrov2016qdo1612.09298,Wang1901.11537}.\\
$\bullet$ Fractional C-P-T symmetry does not necessarily imply fractional spin-statistics of anyons beyond fermions.
For example, the 3+1d Dirac spinor of \eq{CPT-D8-Z4-central} and \eq{eq:projective-linear-representation}
shows that the fermion $\psi$ sits in the projective {rep} of $G_{\phi}$ and carries fractionalized C-P-T quantum numbers of $G_{\phi}$, 
but $\psi$ sits in the (anti)linear {rep} of $\tilde G_{\psi}$.
The $\psi$ does not have anyonic statistics, but only has fermionic statistics (spin $s=1/2$, but still fractionalized with respect to a bosonic integer spin).\\
$\bullet$ Vice versa, fractional spin-statistics of anyons do not imply a fractional C-P-T symmetry, 
because intrinsic topological orders (that give rise to anyons) do not necessarily require any global symmetry.
\\
\noindent
$\bullet$ The  {\bf  spin-statistics theorem} colloquially says the \emph{self-braiding statistics} of an excitation
can be deformed to the \emph{mutual-braiding statistics} between two (or more) excitations, illustrated by Dirac Belt and Feynman Plate tricks \cite{FeynmanWeinberg}. 
Thus this theorem reveals the {\bf topological properties of matter}:
the topological links of world-trajectories of (semiclassical or entangled quantum) matter excitations inside the spacetime manifold.
\\
\noindent
$\bullet$ The  {\bf  CPT or CRT theorem} colloquially says 
that our physical laws are also obeyed by a CRT image of our universe.
Thus this theorem reveals the {\bf topological properties of spacetime}, 
the disconnected components of the spacetime symmetry groups, 
and how the matter-antimatter are transformed under those discrete symmetries.
\\
\noindent
$\bullet$ We propose that the relation between the {spin-statistics theorem and the CPT theorem} may also shed light on the
relation between the  {\bf fractional spin-statistics} and the {\bf fractionalized C-P-R-T} structure. 
{Follow the promise of the fractional spin-statistics studies in the past 
decades \cite{Wilczek1990BookFractionalstatisticsanyonsuperconductivity, Nayak0707.1889}, 
we anticipate that the fractional C-P-R-T topic 
presented here 
will also offer various future applications, both relativistic or nonrelativistic, in high-energy physics or quantum material systems.}\\

\noindent
\emph{Acknowledgments} ---
JW thanks Pierre Deligne, Dan Freed, Jun Hou Fung, Ryan Thorngren, and is especially grateful 
to Pavel Putrov, Zheyan Wan, Shing-Tung Yau, Yi-Zhuang You, Yunqin Zheng, and Martin Zirnbauer for helpful comments. 
JW appreciates Professor 
Yau for persistently raising the question: 
``Can C-P-T symmetries be fractionalized more than $\Z_2$-involutions?''
JW also thanks Abhishodh Prakash for the past collaborations on the fractionalization of time-reversal T symmetry in \cite{PrakashJW2011.12320, PrakashJW2011.13921}.
This work is supported by 
Harvard University CMSA.

\newpage

 \tableofcontents
 
 \onecolumngrid

\bibliography{CPTbib.bib}

\end{document}